\documentstyle[epsfig,11pt]{article}
\renewcommand{\theequation}{\arabic{equation}}
\def\lsim{\mathrel{\rlap{
\lower4pt\hbox{\hskip-3pt$\sim$}}
    \raise1pt\hbox{$<$}}}     
\def\gsim{\mathrel{\rlap{
\lower4pt\hbox{\hskip1pt$\sim$}}
    \raise1pt\hbox{$>$}}}     

\newcommand{\beq}{\begin{equation}}
\newcommand{\eeq}{\end{equation}}
\newcommand{\be}{\begin{eqnarray}}
\newcommand{\ee}{\end{eqnarray}}

\def\np{Nucl. Phys.}
\def\pr{Phys. Rev.}
\def\prl{Phys. Rev. Lett.}
\def\pl{Phys. Lett.}
\def\del{\partial}
\def\Tr{{\rm Tr}}
\def\L{{\cal L}}
\def\Linv{{\cal L}_{inv}}
\def\Lsb{{\cal L}_{sb}}
\def\la{\langle}
\def\ra{\rangle}
\def\der{\mbox{d}}

\begin{document}
\setcounter{equation}{0}

\begin{titlepage}
\pagestyle{empty}
\vspace{1.0in}
\begin{flushright}
February 1996
\end{flushright}
\vspace{1.0in}
\begin{center}
\begin{Large}
{\bf{FROM CHIRAL LAGRANGIANS}}\\
{{\bf TO LANDAU FERMI LIQUID THEORY}}
\vskip 0.15cm
{\bf OF NUCLEAR MATTER}
\end{Large}
\vskip 0.7in
{\large Bengt Friman}
\vskip 0.15cm
{
{\it GSI, Planckstr. 1\\
D-64291 Darmstadt, Germany \&}}

{\it Institut f\"ur Kernphysik, TH Darmstadt\\
D-64289 Darmstadt, Germany}
\vskip 0.2cm
and
\vskip 0.2cm
{\large Mannque Rho}
\vskip 0.15cm
{
{\it CEA Saclay, Service de Physique Th\'eorique\\
91191 Gif-sur-Yvette Cedex, France}}
\end{center}
\vspace{1.2cm}

\begin{abstract}

A simple relation between the effective parameters of chiral 
Lagrangians in medium as predicted by BR scaling
and Landau Fermi liquid parameters is derived. This provides a link
between an effective theory of QCD at mean-field level
and many-body theory of nuclear matter. It connects in particular
the scaling vector-meson mass probed by dileptons produced 
in heavy-ion collisions (e.g., CERES of CERN-SPS)
to the scaling nucleon-mass relevant for low-energy
spectroscopic properties, e.g., the nuclear gyromagnetic ratios $\delta g_l$
and the effective axial-vector constant $g_A^\star$.
\end{abstract}
\end{titlepage}

\section{Introduction}
\indent

In recent publications \cite{lkb95}, Li, Ko and Brown showed that the dilepton
production data of CERES \cite{CERES} and HELIOS-3 \cite{HELIO}
can be simply and quantitatively understood if the mass
of the vector mesons $\rho$ and $\omega$ scales in dense and/or hot medium
according to the scaling (BR scaling)
proposed by Brown and Rho \cite{br91}. That
the vector mesons ``shed" their masses as the density (or temperature)
of the matter increases is expected in an intuitive interpretation of
the interplay of the condensation of quark-antiquark pairs
and the dynamical generation of light-quark hadron masses
and is in fact corroborated by QCD sum rules \cite{sumrules,Jin1}
and model calculations \cite{model}. Thus, the dilepton data are consistent
with the most conspicuous prediction of BR scaling
\footnote{This of course does not exclude other
explanations based on different dynamical schemes
with different Lagrangians such as, e.g., that in \cite{wambach}. It should,
however, be borne in mind that such explanations are not necessarily
{\it alternative} ones; they may in fact be overlapping to a varying degree
in physics.}.
The proposal of \cite{br91}, however, goes further than this and
makes a statement on the relation between the scaling of
meson masses and that of baryon masses:
\be
\frac{m_M^\star}{m_M}\approx \sqrt{\frac{g_A}{g_A^\star}}
\frac{m_B^\star}{m_B}\approx \frac{f_\pi^\star}{f_\pi}\equiv
\Phi (\rho)\label{br}
\ee
where the subscript $M$ stands for light-quark non-Goldstone mesons,
$B$ for light-quark baryons, $g_A$ the axial-current coupling
constant and $f_\pi$ the pion decay constant.
The star denotes an in-medium quantity. (Although temperature
effects can also be discussed in a similar way, we will
be primarily interested in density effects in this paper.)

Two important questions remained unanswered in these developments:
Firstly, is there evidence that the baryon mass scaling and the
meson mass scaling are related as implied by the chiral Lagrangian?
Secondly, we know from the Walecka model of nuclear matter \cite{waleckamodel}
that the ``scalar mass" of the nucleon drops as a function of density
and that this reduction of the nucleon mass has significant consequences on
nuclear spectroscopy and the static properties of nuclei. The question is:
Is BR scaling related to the ``conventional" mechanism for the reduction of
the nucleon mass in nuclear matter and if so, how does it manifest itself in
low-energy nuclear properties?

The purpose of this paper is to show, based on
recent work \cite{FR,br95}, that the connection between the
meson and baryon scalings can be
made using the Landau-Migdal theory of nuclei and nuclear matter.
A similar attempt was recently made by Brown \cite{GEB95}. Our starting
point is the effective chiral Lagrangian used in \cite{br91} where
the scale anomaly of QCD is incorporated
and baryons arise as skyrmions. This theory is mapped onto an
effective meson-baryon chiral Lagrangian. We establish the relation between
chiral and Walecka mean fields in medium as suggested in \cite{br95} and then
invoke the Galilei invariance argument of Landau, which relates the nucleon
effective mass to the Landau Fermi liquid parameters. Thus, we establish a
relation between the parameters in eq.~(\ref{br}) and the Landau parameters. We
discuss how this relation can be tested with the effective $g_A^\star$ and
the gyromagnetic ratios $\delta g_l$ in nuclear matter. This then supplies
an intriguing -- and hitherto undiscovered -- relation between the scaled
masses, which may be reflected in the spectrum of dileptons produced in
relativistic heavy-ion collisions, and low-energy spectroscopic information,
$g_A^\star$ and $\delta g_l$. It also supplies an albeit indirect and poorly
understood connection between quantities figuring in chiral Lagrangians of
QCD and those appearing in familiar many-body theory. This
connection indicates that low-energy effective theories can provide
important insight necessary to understand ultrarelativistic
heavy-ion reactions in which QCD variables are relevant.

In order to avoid unnecessary complications we shall use the
nonrelativistic approach to Landau Fermi liquid theory, referring to
results obtained in the relativistic formulation \cite{baymchin,BWBS}
where appropriate.
The latter approach is briefly discussed in the Appendix.

\section{BR Scaling in Chiral Lagrangians}
\indent

The BR scaling relation (\ref{br}) that relates the dropping of light-quark
non-Goldstone-boson masses to that of the nucleon mass which in
turn is related to that of the pion decay constant was first derived
by incorporating the trace anomaly of QCD
into an effective chiral Lagrangian. The basic idea can be summarized as
follows.  We wish to write an effective chiral Lagrangian which at
mean-field level reproduces the quantum trace anomaly while including
higher chiral order effects relevant for nuclear dynamics. To do this,
we write the effective Lagrangian in two parts
\be
\L=\Linv +\Lsb \label{twopieces}
\ee
where $\Linv$ is the scale-invariant part and $\Lsb$ the
scale-breaking part of the effective Lagrangian. We introduce the {\it
chiral-singlet} scalar field $\chi$, as an interpolating field for
$\Tr\ G^2$, \be \theta_\mu^\mu=\frac{\beta (g)}{2g} \Tr\,
G_{\mu\nu}G^{\mu\nu}\equiv \chi^4, \ee where we have dropped the quark
mass term (here we consider the chiral limit). The simplest possible 
invariant piece of the Lagrangian then takes the form
\be
\Linv=\frac{f_\pi}{4} \left(\frac{\chi^2}{\chi_0^2}\right)\Tr\, (\del_\mu U
\del^\mu U^\dagger) + \frac{1}{32  g^2}\Tr\, [U^\dagger\del_\mu U,
U^\dagger\del_\nu U]^2 +\cdots\label{effectiveL}
\ee
where $\chi_0$ is a number which we define to be the expectation value
of $\chi$ in matter-free vacuum and
the ellipsis stands for other-scale invariant terms including
the kinetic energy term for the $\chi$ field. Note that this is the simplest
possible form based on the most economical assumption.  One could perhaps write
much more complicated and yet scale-invariant forms using the same
set of fields but invoking different assumptions, and thus obtain a 
different type of scaling. Experiments will tell us which one is the right
form.

As for the scale-breaking term $\Lsb$, we assume that it
contains just the terms needed to reproduce the full trace anomaly. We 
add other scale-invariant terms representing higher chiral order terms to
assure the correct vacuum potential which we shall call $V(\chi,U)$.
Fortunately all we need to know about the potential $V$ is that it
contains a source for the $\chi$ mass term and that, for a given density, it
attains its minimum at $\chi^\star=\la \chi \ra^\star$ in
the sense of the Coleman-Weinberg mechanism \cite{coleman}. (We will return
later to what this quantity $\chi^\star$ represents physically.)

The fact that the vacuum expectation value is obtained by minimizing the
potential, which contains a scale-breaking term, implies that we are
treating the breaking of the scale invariance as a spontaneous symmetry
breaking. It is well-known that the spontaneous breaking of the scale
symmetry occurs only if it is explicitly broken, since otherwise the
potential would be flat \cite{zumino}. Given the ground state
characterized by $\chi^\star$ which is fixed by the anomaly, we then 
shift the field in (\ref{twopieces}) 
\be
\chi (x)=\chi^\prime (x) +\chi^\star.
\ee
After shifting, we still have the scale-invariant and scale-breaking
pieces although the manifest invariance is lost as is the case with {\it all}
spontaneously broken symmetries. The low-energy physics for the scaling
we are interested in is lodged in the former. Since the theory contains two
parameters, $f_\pi$ and $g$, we define
\be
f_\pi^\star&=&f_\pi\frac{\chi^\star}{\chi_0},\nonumber\\
g^\star &=& g.
\ee
The second relation follows since the Skyrme quartic term in (\ref{effectiveL})
is scale-invariant by itself.\footnote{We will argue later that
in the baryon sector
there is an important radiative correction -- absent in the meson sector --
which modifies this scaling behavior.}
This allows us to redefine the parameters that appear in the chiral
Lagrangian in terms of the ``starred" parameters $f_\pi^\star$ and
$g^\star$. Since the KSRF relation \cite{KSRF} is an exact low-energy theorem
as shown by Harada, Kugo and Yamawaki \cite{harada}, it is reasonable to
assume that it holds also in medium. This
leads to
\be
m_V^\star/m_V\approx \frac{f_\pi^\star g^\star}{f_\pi g}\approx
\frac{f_\pi^\star}{f_\pi}
\equiv\Phi (\rho) <1\ \ \ {\rm for}\ \ \rho\neq 0
\ee
where the subscript V stands for $\rho$ or $\omega$ meson.
Similarly the mass of the scalar field is reduced
\be
m_\sigma^\star/m_\sigma\approx \Phi (\rho).
\ee
Here we denote the relevant scalar field by the usual notation
$\sigma$ for reasons given below.

Now in order to find the scaling behavior of the nucleon mass, we use the
fact that the nucleon arises as a soliton (skyrmion) from the effective
chiral Lagrangian as in the free-space. The soliton mass goes like
\be\label{skyrmion}
m_S\sim f_\pi/g.
\ee
If one assumes that by the same token the coupling constant $g$ in the
soliton sector is not modified in the medium, eq.~(\ref{skyrmion}) implies
that the nucleon mass is also proportional to $f_\pi^\star$,
\be
m_N^\star/m_N\sim  \Phi (\rho).
\ee
However there is a caveat to this. When it comes to the
nucleon effective mass, there is one important non-mean-field effect
of short range that is known to be important. This is an
intrinsically quantum effect that cannot be accounted for in low orders of
the chiral expansion, namely the mechanism that quenches the axial-current
coupling constant $g_A$ in nuclear matter.
This effect is closely related to the Landau-Migdal interaction in the
spin-isospin channel $g_0^\prime$ (involving $\Delta$-hole excitations)
as discussed in \cite{brPR95,delta}. The
axial-vector coupling constant of the skyrmion is governed by
coefficient $g$ of the Skyrme quartic term. This implies that
in the baryon sector, the mean-field argument, which is valid in the mesonic
sector, needs to be modified. This is reminiscent of the deviation in the
nucleon electromagnetic form factor from the vector dominance model which
works very well for non-anomalous processes involving mesons.
These two phenomena may be related.

As shown in \cite{br91,mr88}, a more accurate expression, at least
for densities up to $\rho\sim \rho_0$, is\footnote{This was
derived using the scaling behavior of the Skyrme quartic
Lagrangian and the relation between $g_A$ and the coefficient $g$.
Although this relation is justified strictly at the large $N_c$ limit
(where $N_c$ is the number of colors), we think
that it is generic and will emerge in
any chiral model that has the correct symmetries.}
\be
m_N^\star/m_N\approx \Phi (\rho)\sqrt{\frac{g_A^\star}{g_A}} .\label{Nmass}
\ee
This relation will be used later to deduce a formula for $g_A^\star$
in nuclear matter.
Beyond $\rho=\rho_0$, we expect that $g_A^\star$ remains constant
($g_A^\star = 1$) and that $\Phi$ scaling takes over
except near the chiral phase transition at which the coupling constant
$g$ will fall according to the ``vector limit." \cite{brPR95}

\subsection{The meaning of $\chi^\star$}
\indent

The $\chi$ field interpolating as $\chi^4$ for the dimension-4 field
$\Tr G_{\mu\nu}^2$ may be dominated by a scalar glueball field, which
perhaps could be identified with the $f_J (1710)$ seen in lattice
calculations \cite{lattice}. However, for the scaling we are
discussing which is an intrinsically low-energy property, this is too
high in energy scale. In the  
effective Lagrangian (\ref{effectiveL}), such a heavy degree of
freedom should not appear explicitly.  The only reasonable interpretation
is that the $\chi$ field has two components,
\be
\chi=\chi_h + \chi_l\label{sep}
\ee
corresponding to high (h) and low (l) mass excitations, and 
that the high mass (glueball) component $\chi_h$ is
integrated out.  The ``vacuum" expectation value we are interested in
is therefore $\la \chi_l \ra^\star$. The corresponding fluctuation
must interpolate $2\pi$, $4\pi$ etc. excitations as discussed in
\cite{brPR95} and it is this field denoted by $\sigma$ that becomes
the dilaton degenerate with the pion at the chiral phase transition as
suggested by Weinberg's mended symmetry \cite{mended}.  It is also
this component which plays an essential role in the relation between
chiral Lagrangians and the Walecka model \cite{walecka,br95}.
This procedure may also be justified by a phenomenological instanton
model anchored in QCD \cite{bbr96}.

For a more physical interpretation and a detailed discussion on the 
separation (\ref{sep}), see Adami and
Brown \cite{adamibrown}. A somewhat different separation is advocated
by Furnstahl et al. in \cite{walecka}.

\subsection{Baryon chiral Lagrangian}
\indent

In order to make contact with many-body theory of nuclear matter, we
reinterpret the BR scaling in terms of a
baryon chiral Lagrangian in the relativistic baryon
formalism. There is a problem with chiral counting in this formalism\footnote{
As we know from the work of Gasser, Sainio and Svarc \cite{gasser},
the relativistic
formulation of baryon chiral perturbation theory requires a special care
in assuring a correct chiral counting. What we will find below is that in
order to get to the correct formulation from the point of view of Landau
Fermi liquid theory of normal nuclear matter
and making contact with Walecka theory at mean-field
order, it is essential to keep relativistic corrections
from the start. This probably
has to do with the presence of the Fermi sea in the effective chiral
Lagrangian approach. This seems to suggest that the usual chiral counting
valid in free space needs to be modified in medium.} but
our argument will be made at mean-field order as in \cite{br95}.

The Lagrangian contains the usual pionic piece $\L_\pi$, the pion-baryon
interaction $\L_{N\pi}$ and the four-Fermi contact interactions
\be
\L_{4}=\sum_\alpha \frac{C_\alpha^2}{2} (\bar{N}\Gamma_\alpha N)
(\bar{N}\Gamma^\alpha N)
\ee
where the
$\Gamma^\alpha$'s are Lorentz covariant quantities -- including derivatives --
that have the
correct chiral properties. The leading chiral order  four-Fermi contact
interactions relevant for the scaling masses are of the form
\be
\L_{4}^{(\delta)}
=\frac{C_\sigma^2}{2} (\bar{N}N\bar{N}N) -\frac{C_\omega^2}{2}
(\bar{N}\gamma_\mu N\bar{N} \gamma^\mu N).\label{wform}
\ee
As indicated by our choice of notation, the first term can be thought of as
arising when a massive isoscalar scalar meson (say, $\sigma$) is integrated
out and similarly for the second term involving a massive isoscalar vector
meson (say, $\omega$). Consequently, we can make the identification
\be
C_\sigma^2=\frac{g_\sigma^2}{m_\sigma^2},\ \
C_\omega^2=\frac{g_\omega^2}{m_\omega^2}.\label{constant}
\ee
The four-Fermi interaction involving the $\rho$ meson quantum number will be
introduced below, when we consider the  electromagnetic currents.
As is well known \cite{walecka,br95}, the first four-Fermi  interaction in
(\ref{wform}) shifts the nucleon mass in matter,
\be
m_N^\sigma=m_N-C_\sigma^2 \la \bar{N}N\ra. \label{sigmashift}
\ee
In \cite{br95} it was shown that this shifted nucleon mass scales the same
way as the vector and scalar mesons
\be \frac{m_V^\star}{m_V}\approx
\frac{m_\sigma^\star}{m_\sigma}\approx \frac{m_N^\sigma} {m_N}\approx  \Phi
(\rho).\label{universal}
\ee
This relation was referred to in \cite{br91} as ``universal scaling." There
are two points to note here: First as argued in \cite{br95}, the
vector-meson mass scaling applies also to the masses in (\ref{constant}).
Thus, in medium the meson mass should be replaced by $m_{\sigma,
\omega}^\star$. Consequently, the coupling strengths $C_\sigma$ and
$C_\omega$ are density-dependent. Second, the scaling can be understood in
terms of effects due to the four-Fermi interactions, which for nucleons on
the Fermi surface correspond to the
fixed-point interactions of Landau Fermi liquid theory according to Shankar
and Polchinski \cite{shankarpol}. We shall establish a direct connection to
the Landau parameters of the quasiparticle-interaction.

\section{Effective Nucleon Mass \`a la Landau}
\indent

In the Landau-Migdal Fermi liquid theory of nuclear matter
\cite{landau,migdal}, the interaction between two
quasiparticles on the Fermi surface is of the form (neglecting tensor
interactions)
\beq
{\cal F}(\vec{p},\vec{p}^\prime) = F(\cos \theta) + F^\prime (\cos
\theta)(\vec{\tau}\cdot\vec{\tau}^\prime) + G(\cos
\theta)(\vec{\sigma}\cdot\vec{\sigma}^\prime) + G^\prime (\cos \theta)
(\vec{\tau}\cdot\vec{\tau}^\prime)
(\vec{\sigma}\cdot\vec{\sigma}^\prime),
\eeq
where $\theta$ is the angle between $\vec{p}$ and
$\vec{p}^\prime$. The function $F(\cos \theta)$ can be expanded in
Legendre polynomials,
\beq
F(\cos \theta) = \sum_l F_l P_l(\cos \theta),
\eeq
with analogous expansions for the spin- and isospin-dependent
interactions. The coefficients $F_l$ etc. are the Landau Fermi liquid
parameters. Some of the parameters can be related to physical
properties of the system. The relation between the effective mass and
the Landau parameter $F_1$ (eq.~(\ref{mstar})) is crucial for our
discussion.

\begin{figure}[bth]
\center{\epsfig{file=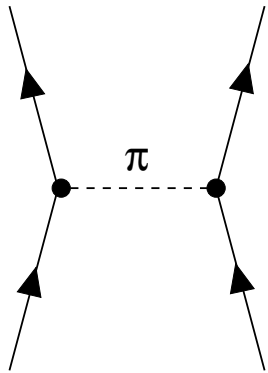,
height=35mm,angle=-90}}
\hfill
\caption{\label{opepfig} The
one-pion-exchange interaction corresponding to the non-local
four-Fermi term in the Lagrangian (\protect{\ref{4fermii}}).}\hfill
\end{figure}

An important point of this paper is that one must distinguish between
the effective mass $m_N^\sigma$, which is of the same form as
Walecka's effective mass, and the Landau effective mass, which
is more directly related to nuclear observables.
To see what the precise relation is, we include the non-local
four-Fermi interaction due to the one-pion exchange term, $L^{(\pi)}_4$,
shown in fig.~\ref{opepfig}.

\begin{figure}[t]
\center{\epsfig{file=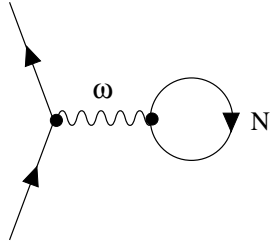,
height=35mm,angle=-90}}
\hfill\caption{\label{omega_hartree} The $\omega$-meson contribution
to the nucleon self energy.}\hfill
\end{figure}

The total four-Fermi interaction that enters in the renormalization-group
flow consideration \`a la Shankar-Polchinski is then the sum
\be
\label{4fermii}
\L_4=\L_4^{(\pi)} +\L_4^{(\delta)}.
\ee
The point here is that the non-local one-pion-exchange term brings
additional contributions to the effective nucleon mass on top of the
universal scaling mass discussed above.
\begin{figure}[t]
\center{\epsfig{file=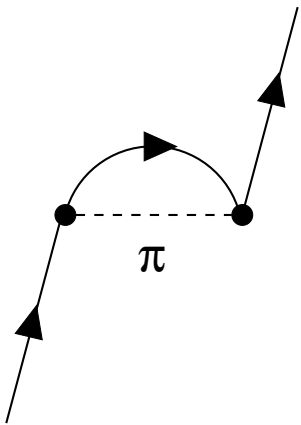,
height=40mm,angle=-90}}
\hfill\caption{\label{fockfig} The one-pion exchange (Fock term)
contribution to the nucleon self energy.}\hfill
\end{figure}
\noindent
We now compute the nucleon effective mass with the chiral Lagrangian and
make contact with the results of Fermi liquid theory \cite{FR}.
We start with the single-nucleon energy in the non-relativistic
approximation\footnote{We treat the scalar and vector fields
self-consistently, as described in the Appendix, and the self-energy from
the pion exchange graph as a perturbation.}
\be
\epsilon (p) =\frac{p^2}{2 m_N^\sigma} +
C_\omega^2\la N^\dagger N\ra +\Sigma_\pi (p)
\label{energy}
\ee
where $\Sigma_\pi (p)$ is the self-energy from the pion exchange graph,
shown in fig.~\ref{fockfig}.
The self-energy contribution from the vector meson (second term on the
right hand side of (\ref{energy})) corresponds to the diagram shown in
fig.~\ref{omega_hartree}.
The Landau effective mass $m_L^\star$ is related to the quasiparticle
velocity at the Fermi surface
\be
\label{velo}
\frac{\der}{\der p} \epsilon (p)|_{p=p_F}=\frac{p_F}{m_L^\star}
= \frac{p_F}{m_N^\sigma} +\frac{\der}{\der p}\Sigma_\pi (p)|_{p=p_F}.
\ee
Using Galilean invariance, Landau \cite{landau} derived a relation between the
effective mass of the quasi-particles and the velocity dependence of the
effective interaction described by the Fermi-liquid parameter $F_1$:
\be
\label{mstar}
\frac{m^\star_L}{m_N} = 1 + \frac{F_1}{3} = (1-\frac{\tilde{F_1}}{3})^{-1},
\ee
where $ \tilde{F_1} = (m_N/m^\star_L) F_1$.
The corresponding relation
for relativistic systems follows from Lorentz invariance and has been
derived by Baym and Chin \cite{baymchin} (see Appendix).

With the four-Fermi interaction (\ref{4fermii}), there are two distinct
velocity-dependent terms in the quasiparticle interaction, namely the
spatial part of the current-current interaction and the exchange (or Fock)
term of the one-pion-exchange. In the nonrelativistic approximation, their
contributions to $\tilde{F_1}$ are ($\tilde{F_1} = \tilde{F_1^\omega} +
\tilde{F_1^\pi}$) 
\be
\label{fomega}
\tilde{F_1^\omega}&=&\frac{m_N}{m_L^\star} F_1^\omega=
-C_\omega^2\frac{2p_F^3}{\pi^2 m_N^\sigma },\\
\tilde{F_1^\pi}&=& -3\frac{m_N}{p_F}\frac{\der}{\der p}\Sigma_\pi (p)|_{p=p_F},
\ee
respectively. The relativistic expression for $\tilde{F_1^\omega}$ is given
in the Appendix.

Using eq.~(\ref{velo}) we find
\be
(\frac{m_L^\star}{m_N})^{-1}
=\frac{m_N}{m_N^\sigma} +\frac{m_N}{p_F}\frac{\der}{\der p}\Sigma_\pi
(p)|_{p=p_F} = 1-\frac 13 \tilde{F_1}\label{LandauM},
\ee
which implies that
\be
\frac{m_N}{m_N^\sigma}=1-\frac 13 \tilde{F}_1^\omega.\label{Phidefined}
\ee
This is an interesting relation between the $\sigma$-nucleon interaction
(eq.~(\ref{sigmashift})) and the $\omega$-nucleon coupling 
(eq.~(\ref{fomega})). The $\omega$-exchange contribution to the Landau
parameter $F_1$ is due to the velocity-dependent part of the
potential, $\sim \vec{p}_1\cdot\vec{p}_2/m_N^2$.
This is an ${\cal O}(p^2)$ term, and consequently suppressed in naive chiral
counting.  Nonetheless it is this chirally non-leading term in the
four-Fermi interaction (\ref{wform}) that appears on the same footing
with the chirally leading terms in the $\omega$ and $\sigma$ tadpole
graphs. This suggests a subtlety in the chiral counting in the
presence of a Fermi sea.

The pion contribution to $F_1$ can be evaluated explicitly \cite{br80}
\be
\frac 13
\tilde{F_1^\pi}=-\frac{3f_{\pi NN}^2m_N}{8\pi^2p_F}[\frac{m_\pi^2+2p_F^2}
{2p_F^2} \ln\frac{m_\pi^2+4p_F^2}{m_\pi^2}-2]\approx -0.153.\label{FockM}
\ee
Here $f_{\pi NN}\approx 1$ is the non-relativistic $\pi$N coupling
constant.
The numerical value of $\tilde{F_1^\pi}$ is obtained at nuclear matter density,
where $p_F\approx 2m_\pi$.

One of the important results of this paper is that
eq.~(\ref{Phidefined}) relates the only unknown
parameter $\tilde{F}_1^\omega$ to the universal scaling factor $\Phi$.
Note that in the absence of the one-pion-exchange interaction -- and
in the nonrelativistic approximation --
$m_N^\sigma$ can be identified with the Landau effective mass
$m_L^\star$. In its presence,
however, the two masses are different due to the pionic Fock term.
We propose to identify the scaling nucleon mass defined in
eq.~(\ref{Nmass}) with the Landau effective mass:
\be\label{mrelation}
m_L^\star=m_N^\star.
\ee
We note that the Landau mass is defined at the Fermi surface, while the
scaling mass refers to a nucleon propagating in a ``vacuum" modified
by the nuclear medium. Although  
the two definitions are closely related, their precise
connection is not understood at present. Nevertheless,
eq.~(\ref{mrelation}) is expected to be a good approximation 
(see also section 5.2).

\section{Nucleon Gyromagnetic Ratios in Nuclei}
\indent

Given the effective Lagrangian with the BR scaling and its relation
to Landau Fermi liquid theory,
how can one describe nuclear magnetic moments and axial charge transitions?
This is an important question
because these nuclear processes are sensitive to both the
scaling properties and exchange currents. Here we consider the
gyromagnetic ratios $g_l^{(p,n)}$ of the proton and the neutron in heavy
nuclei, deferring the issue of the nuclear axial-charge transitions
\cite{pmr93} to a later publication \cite{FR}.
We start with the Fermi liquid theory result for the gyromagnetic 
ratio.\footnote{This quantity has been extensively analyzed in terms of
standard exchange currents and their relations, via vector-current
Ward identities, to nuclear forces \cite{riska}.}

\subsection{Migdal's formula for the gyromagnetic ratio}
\indent

The response to a slowly-varying electromagnetic field of an odd
nucleon with momentum $\vec{p}$ added to a closed Fermi sea
can, in Landau theory, be represented by the current \cite{migdal,BWBS}
\be
\vec{J}= \frac{\vec{p}}{m_N}\left(\frac{1+\tau_3}{2} +\frac 16
\frac{F_1^\prime -F_1}{1+F_1/3} \tau_3\right) \label{qpcurrent}
\ee
where $m_N$ is the nucleon mass in medium-free space.
The long-wavelength limit of the current is not unique. As discussed in
the Appendix, the physically relevant one
corresponds to the limit $q \rightarrow 0, \omega \rightarrow 0$ with
$q/\omega \rightarrow 0$, where $(\omega,q)$ is the four-momentum transfer.
The current (\ref{qpcurrent}) defines the gyromagnetic ratio
\be
g_l=\frac{1+\tau_3}{2} +\delta g_l
\ee
where
\be
\delta g_l=\frac 16 \frac{F_1^\prime -F_1}{1+F_1/3}\tau_3
=\frac 16 (\tilde{F}_1^\prime-\tilde{F}_1)\tau_3.\label{deltalandau}
\ee

\subsection{Chiral Lagrangian prediction}
\indent

\begin{figure}[tbh]
\center{\epsfig{file=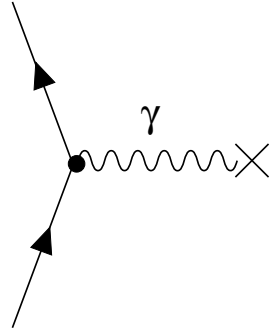,
height=35mm,angle=-90}}
\hfill\caption{\label{sp_current} The single particle current.}\hfill
\end{figure}

\begin{figure}[tbh]
\center{\epsfig{file=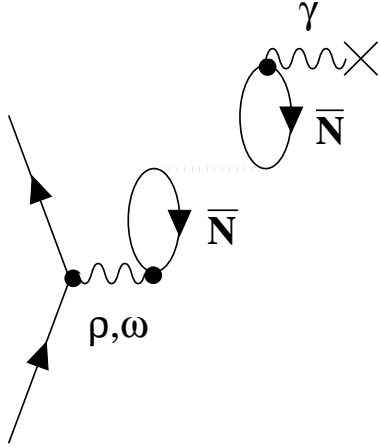,
height=50mm,angle=-90}}
\caption{\label{pol_current} The polarization contribution to the
current. The backward-going lines correspond to negative energy states
(anti-nucleons) while the lines without arrows represent nucleon
states that are blocked by the filled Fermi sea.}
\end{figure}

In this section we compute the gyromagnetic ratio using the chiral 
Lagrangian and demonstrate that Migdal's result (\ref{deltalandau}) is
reproduced. The derivation will be made in terms of Feynman diagrams.
The single-particle current $\vec{J}_1=\vec{p}/m_N^\sigma$ is
given by the diagram shown in fig.~\ref{sp_current}
where the external nucleon lines are dressed by the scalar and vector fields.
Note that it is the universally scaled mass $m_N^\sigma$ that enters,
not the Landau mass. This leads to a gyromagnetic ratio
\be
(g_l)_{sp}=\frac{m_N}{m_N^\sigma}\frac{1+\tau_3}{2}.
\ee
\noindent
The first correction to this is the contribution from short-ranged high-energy
isoscalar vibrations as depicted in fig.~\ref{pol_current},
with the exchanged particle being an $\omega$ meson.
This contribution has been computed by several authors \cite{matsui,suzuki}.
In the nonrelativistic approximation one finds
\be
g_l^\omega=-\frac 16 C_\omega^2\frac{2p_F^3}{\pi^2}
\frac{1}{m_N^\sigma}=\frac 16 \tilde{F}_1^\omega.\label{deltaomega}
\ee
Now using (\ref{Phidefined}), we obtain the second principal result of
this paper,
\be
g_l^\omega=\frac 16 \tilde{F}_1^\omega= \frac 12
(1-\Phi (\rho)^{-1}).\label{f1phi}
\ee
The corresponding contribution with a $\rho$ exchange in the graph
yields an isovector term
\be
g_l^\rho=-\frac 16 C_\rho^2 \frac{2p_F^3}{\pi^2}\frac{1}{m_N^\sigma}
\tau_3=\frac 16 (\tilde{F}_1^\rho)^\prime \tau_3 \label{deltarho}
\ee
where the constant $C_\rho$ is the coupling strength of the four-Fermi
interaction
\be
\delta \L=-\frac{C_\rho^2}{2} (\bar{N}\gamma_\mu \tau^a N
\bar{N}\gamma^\mu \tau^a N).
\ee
In analogy with the isoscalar channel, we may consider this as arising
when the $\rho$ is integrated out from the Lagrangian, and consequently
identify
\be
C_\rho^2=g_\rho^2/m_\rho^2.
\ee
Again in medium, $m_\rho$ should be replaced by $m_\rho^\star$.
The results (\ref{deltaomega}) and (\ref{deltarho})
can be interpreted in the language of chiral perturbation theory
as arising from four-Fermi interaction counterterms in
the presence of electromagnetic
field, with the counter terms saturated by the $\omega$ and $\rho$ mesons
respectively (see eq.~(92) of \cite{pmr95}).

\begin{figure}[tbh]
\center{\epsfig{file=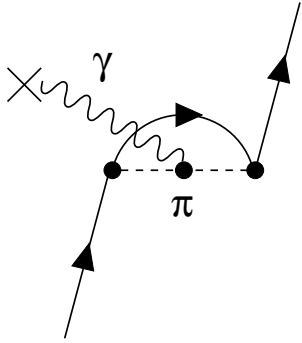,
height=40mm,angle=-90}}
\hfill\caption{\label{exchcurr} The pion-exchange current contribution
to the nucleon current in matter.}\hfill
\end{figure}

The next correction is the pionic exchange current (known as Miyazawa term,
see fig.~\ref{exchcurr}) which yields \cite{br80}
\be
g_l^\pi = \frac 16 ((\tilde{F}_1^\pi)^\prime -\tilde{F}_1^\pi)\tau_3
=-\frac 29 \tilde{F}_1^\pi \tau_3,\label{glpi}
\ee
where the last equality follows from $(\tilde{F}_1^\pi)^\prime = -(1/3)
\tilde{F}_1^\pi$.
Thus, the sum of all contributions is
\be
g_l&=& \frac{m_N}{m_N^\sigma}\frac{1+\tau_3}{2} +\frac 16
(\tilde{F}_1^\omega +(\tilde{F}_1^\rho)^\prime \tau_3)
+ \frac 16 ((\tilde{F}_1^\pi)^\prime -\tilde{F}_1^\pi)\tau_3\nonumber\\
&=&
\frac{1+\tau_3}{2} + \frac 16 (\tilde{F}_1^\prime -\tilde{F}_1)\tau_3
\label{pred}
\ee
where eq.~(\ref{Phidefined}) was used with
\be
\tilde{F}_1 &=& \tilde{F}_1^\omega+\tilde{F}_1^\pi,\label{ftilde}\\
\tilde{F}_1^\prime &=& (\tilde{F}_1^\pi)^\prime +(\tilde{F}_1^\rho)^\prime.
\label{fptilde}
\ee
Thus, using our chiral Lagrangian we reproduce the Fermi-liquid theory result
for $\delta g_l$ (\ref{deltalandau})
\be
\delta g_l =\frac 16 (\tilde{F}_1^\prime-\tilde{F}_1)\tau_3
\label{deltachiral}
\ee
with $\tilde{F}$ and $\tilde{F}^\prime$ in the theory given entirely by
(\ref{ftilde}) and (\ref{fptilde}), respectively.
Equation (\ref{pred}) shows that the isoscalar gyromagnetic ratio is
not renormalized by the medium 
(other than binding effect implicit in the matrix
elements) while the isovector one is. 
{\it It should be emphasized that contrary to naive expectations,
BR scaling is not in conflict with the observed nuclear
magnetic moments.} We will show below that the theory agrees
quantitatively with experimental data.

\section{Test of the Scaling}
\subsection{QCD sum rules}
\indent

It is possible to extract the scaling factor $\Phi (\rho)$ from
QCD sum rules -- as well as from an in-medium Gell-Mann-Oakes-Renner
relation \cite{br95} -- and compare with our theory.
In particular, the key information is available from
the calculations of the masses of the $\rho$ meson \cite{sumrules,Jin1}
and the nucleon \cite{CFG,Jin2}
in medium. In their recent
work, Jin and collaborators find (for $\rho=\rho_0$) \cite{Jin1,Jin2}
\be
\frac{m_\rho^\star}{m_\rho}&=& 0.78\pm 0.08,\\
\frac{m_N^\star}{m_N}&=& 0.67\pm 0.05.\label{Jin}
\ee
We identify the $\rho$-meson scaling with the universal
scaling factor,
\be
\Phi (\rho_0)=0.78.\label{sigmaM}
\ee
This is tantalizingly close to the result that follows from the
GMOR relation in medium \cite{brPR95,CFG2}
\be
\Phi^2 (\rho_0) \approx
\frac{{m_\pi^\star}^2}{m_\pi^2} (1-\frac{\Sigma_{\pi N}\,\rho_0}
{f_\pi^2 m_\pi^2}+\cdots)\approx 0.6,\label{GMOR}
\ee
where the pion-nucleon sigma term $\Sigma_{\pi N}\approx 45$ MeV is used.
\subsection{Prediction}
\indent

Our theory has only one quantity that is not fixed by the theory, namely
the scaling factor $\Phi (\rho)$ ($\tilde{F}_1^\pi$ is of course
fixed for any density by the chiral Lagrangian.).
Since this is given by QCD sum rules
for $\rho=\rho_0$, we use this information to make quantitative prediction.

The first quantity is the Landau effective mass of the nucleon (\ref{LandauM}),
\be
\frac{m_N^\star}{m_N}&=&\Phi\left(1 + \frac 13 F^\pi_1\right) \\
&=& \left(\Phi^{-1}-\frac 13 \tilde{F}_1^\pi\right)^{-1}
=(1/0.78 +0.153)^{-1}=0.69(7)\label{L}
\ee
where we used (\ref{FockM}) and (\ref{sigmaM}).
The agreement with the QCD sum-rule result (\ref{Jin}) is both surprising and
intriguing since as mentioned above, the Landau mass is ``measured" at
the Fermi momentum $p=p_f$ while
the QCD sum-rule mass is defined in the rest frame, so the direct connection
remains to be established.

The next quantity of interest is the axial-vector coupling constant
in medium, $g_A^\star$, which can be obtained from the Landau mass
(\ref{LandauM}) and the chiral mass (\ref{Nmass}) as
\be
\frac{g_A^\star}{g_A}=\left(1+\frac 13 F^\pi_1\right)^2=\left(1-\frac 13
\Phi \tilde{F}_1^\pi\right)^{-2},
\ee
which at $\rho=\rho_0$ gives
\be
g_A^\star=1.0(0).
\ee
This agrees well with the observations in heavy nuclei \cite{gAdata}.
Again this is an intriguing result. While it is not understood how this
relation is related to the old one in terms of the Landau-Migdal parameter
$g_0^\prime$ in $NN\leftrightarrow N\Delta$ channel \cite{delta},
it is clearly a short-distance effect in the ``pionic channel"
involving the factor $\Phi$. This supports the argument \cite{pmr93}
that the renormalization of the axial-vector coupling constant in
medium cannot be described in low-order chiral perturbation theory.

Finally, the correction to the single-particle gyromagnetic
ratio can be rewritten as
\be
\delta g_l=\frac 49\left[\Phi^{-1} -1 -\frac 12 \tilde{F}_1^\pi\right]\tau_3
\ee
where we  have used (\ref{glpi}) and the nonet relation 
$C_\rho^2=C_\omega^2/9$.
At $\rho=\rho_0$, we find
\be
\delta g_l=0.22(7)\tau_3.\label{deltapred}
\ee
This is in agreement with the result \cite{exp} for protons extracted
from the
dipole sum rule in $^{209}$Bi using the Fujita-Hirata relation \cite{FH}:
\be
\delta g_l^{proton}=\kappa/2 = 0.23\pm 0.03.
\ee
Here $\kappa$ is the enhancement factor in the giant dipole sum rule.
Given that this is extracted from the sum rule in the
giant dipole resonance region, this is a bulk property, so our theory
is directly relevant.

Direct comparison with magnetic moment measurements is difficult
since BR scaling is expected to quench the tensor force which is crucial
for the calculation of contributions from high-excitation states
needed to extract the $\delta g_l$. Calculations
with this effect taken into account are not available at present.
Modulo this caveat, our prediction (\ref{deltapred}) compares well
with Yamazaki's analysis \cite{yamazaki} of magnetic moments
in the $^{208}$Pb region
\be
\delta g_l^{proton} &\approx& 0.33,\nonumber\\
\delta g_l^{neutron}&\approx& -0.22
\ee
and also with the result of Arima et al. \cite{arima,yamazaki}
\be
\delta g_l\approx 0.25\tau_3.
\ee

\subsection*{Acknowledgments}

We dedicate this article to Gerry Brown on the occasion of his 70th birthday.
Much of the ideas discussed here came from discussions with him
and were put into a precise form while one of the authors (MR) was
visiting GSI
as a ``Humboldtpreistr\"ager." MR wishes to thank Wolfgang N\"orenberg
and the members of the Theory Group at GSI for hospitality
and the Humboldt Foundation for support.
\newpage
\subsection*{Appendix}
\renewcommand{\theequation}{A.\arabic{equation}}
\setcounter{equation}{0}
\indent

In this appendix, we collect some results of Landau Fermi liquid theory for
relativistic systems. The results are illustrated for a system with scalar and
vector interactions, where the self-consistent mean-field approximation is
tractable. In this approximation, the model is identical to the
Walecka model as noted in \cite{br95}.
To obtain the nonrelativistic results given in the main text,
the long-range nonlocal pion exchange contribution has to be introduced as
a perturbative correction.

The Landau effective mass is related to the quasiparticle velocity at the
Fermi surface, i.e.,
\beq
v_F = \frac{\der \epsilon (p)}{\der p}|_{p=p_F} = \frac{p_F}{m^\star_L}.
\eeq
As shown by Baym and Chin \cite{baymchin}, Lorentz invariance implies a
relation between the effective mass and the quasiparticle interaction,
analogous to the the one derived by Landau for non-relativistic systems
\beq
\label{rmstar}
\frac{m^\star_L}{\mu} = 1 + \frac{F_1}{3} = (1 - \frac{\tilde{F}_1}{3})^{-1}
\eeq
where $\tilde{F}_1 = (\mu/m^\star_L) F_1$ and $\mu$ is the baryon
chemical potential. In the non-relativistic limit $\mu = m_N$, and Landau's
result (\ref{mstar}) is recovered.

In the mean-field approximation to the scalar-vector model, the
single-particle energy is $\epsilon(p) = \sqrt{p^2+(m_N^\sigma)^2} +
C_\omega^2 \la N^\dagger N \ra$, where $m_N^\sigma$ is determined by solving
equation (\ref{sigmashift}) and the last term is due to the time
component of the vector interaction. By computing the quasiparticle velocity
at the Fermi surface, one finds the Landau effective mass in this model,
$m_L^\star = \sqrt{p_F^2+(m_N^\sigma)^2}$.

On the other hand, the only velocity-dependent contribution to the
quasiparticle interaction is due to the spatial part of the vector
interaction. One finds \cite{matsui}
\beq
\tilde{F_1^\omega} = -C_\omega^2
\frac{2p_F^3}{\pi^2 \sqrt{p_F^2+(m_N^\sigma)^2}},
\eeq
i.e., eq.~(\ref{fomega}) with the
appropriate $m_L^\star$. It is easy to check that this model satisfies the
relativistic effective mass relation (\ref{rmstar}).

In an interacting system, the current $\vec{J}_p$ carried by a
quasiparticle with momentum $\vec{p}$ is not necessarily equal to its
velocity $\vec{v}_p$. The difference between the quasiparticle
velocity and the current is the so-called backflow current \cite{PN}.
Diagrammatically the backflow current is associated with a
polarization of the nuclear medium of the particle-hole type. The
particle-hole pairs contribute to the current in the long-wave-length limit
for ${\omega}/{q} \rightarrow 0$ and vanish in the opposite limit
${q}/{\omega} \rightarrow 0$. In the former limit the current is
determined by the Ward identities to be equal to the quasiparticle velocity
\cite{BAHSY}. Physically, this limit corresponds to a localized
quasiparticle excitation \cite{PN}, while the gyromagnetic ratio corresponds
to the opposite limit, i.e., to
a homogeneous quasiparticle excitation\footnote{Another argument for why
this is the physically relevant limit is that it guarantees that
the total charge of the system remains constant \cite{BWBS}.}.

The contribution of the particle-hole pairs to the current may be computed
within Landau Fermi-liquid theory (see e.g. \cite{migdal,PN,BAHSY}).
One then finds the total current in a relativistic system, e.g. by
subtracting the particle-hole contribution from the quasiparticle velocity,
\be
\vec{J} &=& \frac{\vec{p}}{m_L^\star}\left((1 + \frac{F_1}{3})
\frac{1 + \tau_3}{2} + \frac 16(F_1^\prime - F_1)
\tau_3\right)\\ \nonumber
&=&\frac{\vec{p}}{\mu}\left(\frac{1 + \tau_3}{2} + \frac 16
\frac{F_1^\prime - F_1}{1 + F_1/3}\tau_3\right).
\ee
This is the relativistic generalization of (\ref{qpcurrent}).

In the scalar-vector model, the quasiparticle velocity is $v_F=p_F/m_L^\star=
p_F/\sqrt{p_F^2+(m_N^\sigma)^2}$. The isoscalar part of the current is
renormalized by the velocity-dependent part of the interaction $F_1$, which
corresponds to the spatial part of the vector interaction. Diagrammatically,
this contribution to the current corresponds to summing up the diagrams
shown in fig.~\ref{pol_current}.  In this model, the isovector part of the
interaction vanishes, so $F_1^\prime = 0$ and the isovector current is
equal to the quasiparticle velocity.  A vector-isovector interaction,
like that generated by the exchange of a $\rho$ meson, as well as
the pion exchange gives rise
to a non-vanishing $F_1^\prime$.

\end{document}